\documentclass[showpacs,12pt]{revtex4}
\usepackage{graphicx}
\usepackage{bm}
\usepackage{multirow}

\begin{document}

\title{On the spin-isospin decomposition of nuclear symmetry energy}

\author{Wenmei Guo$^{a,b}$, M. Colonna$^{a}$, V.
Greco$^{a,c}$, U. Lombardo$^{a}$
\footnote{Corresponding author at: Laboratori Nazionali del
Sud(INFN), via S. Sofia 62, 95123 Catania, Italy, phone:
+39\ 095\ 542\ 277, fax:
+39\ 095\ 71\ 41\ 815, email: lombardo@lns.infn.it. }, H. J. Schulze$^{d}$}
\affiliation{\mbox{$^{a}$}Laboratori Nazionali del Sud(INFN), via S.Sofia 62, 95123 Catania, Italy, \\
\mbox{$^{b}$}Institute of Theoretical Physics, Shanxi University, 030006 Taiyuan, China, \\
\mbox{$^{c}$}Dipartimento di Fisica e Astronomia, via S.Sofia 64, 95123 Catania, Italy, \\
\mbox{$^{d}$}INFN, Sezione di Catania, via S.Sofia 62, 95123 Catania, Italy}

\date{\today}

\begin{abstract}
The decomposition of nuclear symmetry energy into spin and isospin
components is discussed to elucidate the underlying properties of
the NN bare interaction. This investigation was carried out in the
framework of the Brueckner-Hartree-Fock theory of asymmetric
nuclear matter with consistent two and three body forces. It is
shown the interplay among the various two body channels in terms
of isospin singlet and triplet components as well as spin singlet and
triplet ones. The broad range of baryon densities enables  to
study the effects of three body force  moving from low to high
densities.
\end{abstract}

\maketitle

\section{Introduction}
Over the last two decades the nuclear symmetry energy has been one
of most studied observables in nuclear physics for the important
role it plays in the study of the spectroscopy of nuclei,
heavy-ion collisions (HIC) and nuclear astrophysics (for a review
see Ref.\cite{bao}). The symmetry energy is the response of
symmetric nuclear matter (SNM) to a small neutron-to-proton unbalance
(we assume $\beta=(N-Z)/A >0$) and it is the main property of
asymmetric nuclear matter (ANM). In the framework of the
Brueckner-Hartree-Fock (BHF) the energy per particle displays, as
shown in Fig.1, the well known linear $\beta^2$-dependence within
a broad range of nuclear-matter densities\cite{bomb}.
\begin{figure}[htb]
  \centering
  \includegraphics[angle=0,width=12cm]{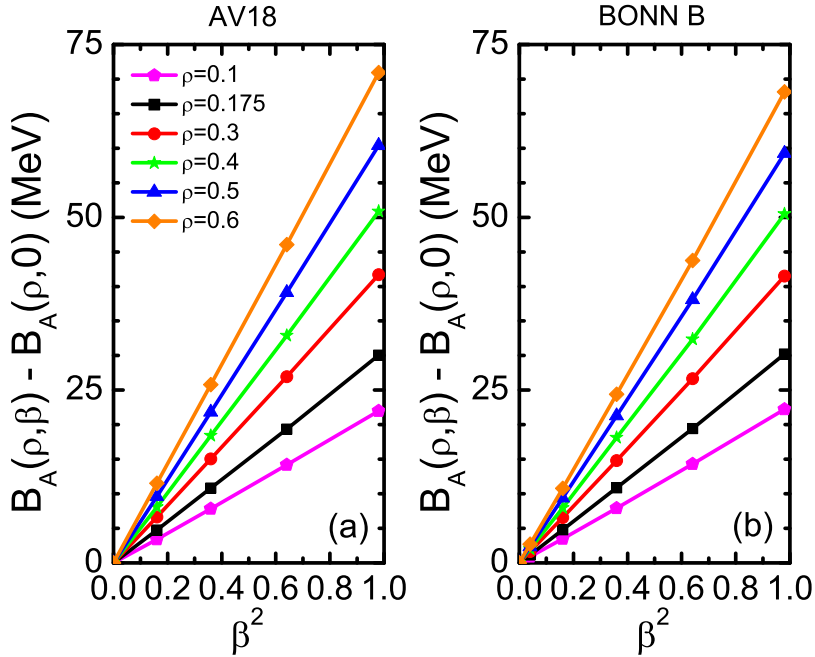}
  \caption{Energy per particle in ANM vs. $\beta^2$ from BHF approximation
  with two different realistic two body forces: AV18 (left) and BONN B (right).}\label{anm2}
\end{figure}
The deviation due to the kinetic part is negligible. Such a
behavior justifies the calculation of the symmetry energy as
difference of binding energies between pure neutron matter (PNM) and symmetric nuclear
matter. Whereas this estimate is validated
by the empirical nuclear mass law, it is would not be so for
neutron stars, whenever the $\beta^2$ linearity was lost. In that
case the symmetry energy should be determined by a small isospin
deviation from the PNM state. From the point of view of the
nucleon-nucleon interaction, the $\beta^2$ linearity seems to be
in conflict with the breakdown of rotational invariance in isospin
space when moving from the bare interaction to the effective
interaction $\mathcal{F}$. Considering the isospin shift of the
single-particle potential $\Delta u_{\tau} =
u_{\tau}(\beta)-u_{\tau}(0)$, it is easily proved that
\begin{equation}
\Delta u_n -\Delta u_p  \,=\, \frac{1}{2}\rho
[\mathcal{F}_{nn}+\mathcal{F}_{pp}-2\mathcal{F}_{np}] \beta \cdot
\end{equation}
The difference $\mathcal{F}_{\tau\tau}-\mathcal{F}_{\tau\tau'}$
($\tau\neq\tau'$) determines the Landau-Migdal parameter
$F'_{\tau}\,$\cite{mig}.Despite  $F'_{n}\neq F'_{p}$ in BHF
approximation, the sum is almost constant at any value of symmetry
parameter for $\rho=\rho_n+\rho_p$ constant\cite{umbe}. This is
also true for any component of  the $E_{sym}$ expansion in two
body channels. In the next section we will discuss the individual
spin-isospin contributions to the symmetry potential energy which
enter in the decomposition
\begin{equation}
U_{sym} \,=\, \sum_{ST} (U^{ST}_{PNM}
-U^{ST}_{SNM} ),
\end{equation}
where $S$ is total spin, $T$ is total isospin, and the z-projection of isospin $T_z$ is dropped out according to the
preceding discussion.

\section{Numerical results}
The spin and isospin decomposition of the the symmetry energy
potential has been calculated in the framework of the BHF
approximation. Two versions two and three body force (2BF and 3BF) have been
employed: Argonne V18 plus consistent meson-exchange
3BF\cite{3bf1} and Bonn B plus consistent meson-exchange
3BF\cite{3bf2}. In Fig.2 it is reported the isospin shift
\begin{equation}
U^T(\rho,\beta) - U^T({\rho,0}) = \sum_S [ U^{ST}(\rho,\beta)
- U^{ST}({\rho,0})]
\end{equation}
with only two body force Bonn B.
\begin{figure}[htb]
\centering
 \includegraphics[angle=0,width=12cm]{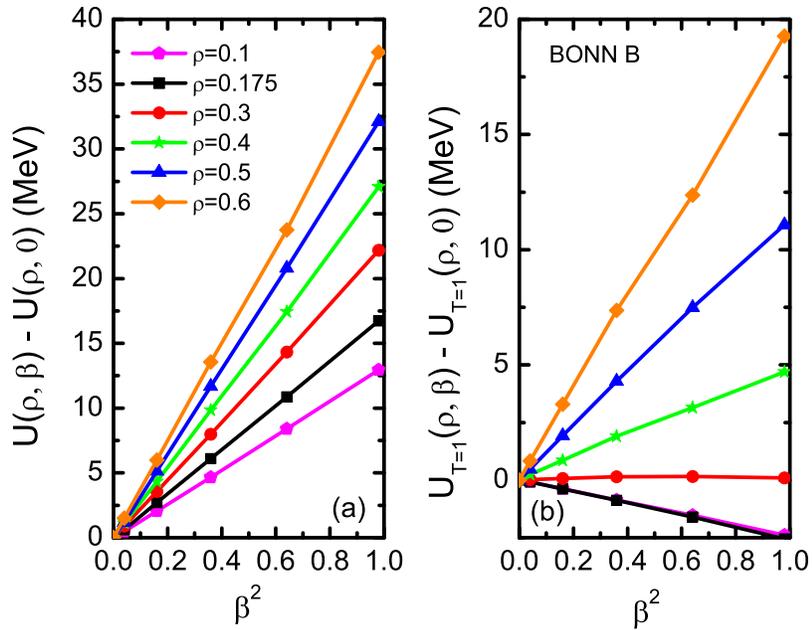}
 \caption{Total potential energy per particle in ANM vs. $\beta^2$
(left)and isospin triplet contribution (right).}\label{t2}
\end{figure}
Due to the simple $\beta^2$ law, one can calculate the total
isospin contribution to the symmetry potential energy from the value of PNM,
namely $\beta=1$. It is seen  around the saturation density that
the isospin-singlet term yields by far the largest contribution to
the symmetry energy  whereas the isospin-triplet is negligible.
The reason is that at low density the G-matrix still keeps the
rotational invariance in isospin space of the bare nucleon-nucleon
interaction, so that the isospin-triplet contribution disappears
($\mathcal{G}^1$ independent of the z-projection $T_z$) as it can be
realized from Eq.1 expressed in term of total isospin
\begin{equation}
\Delta u_n -\Delta u_p  \,=\, \frac{1}{2}\rho
[\mathcal{G}^1_{nn}+\mathcal{G}^1_{pp}-2\mathcal{G}^1_{np}-2\mathcal{G}^0_{np}]
\beta ,
\end{equation}
here
\begin{equation}
\mathcal{G}_{\tau\tau'}^T=\sum_S\mathcal{G}_{\tau\tau'}^{ST},
\end{equation}
and the effective interaction $\mathcal{G}^T(T=0, 1)$ is the G-matrix $G^T$ in the Brueckner theory.
\begin{table}
\renewcommand{\arraystretch}{1.5}
\begin{center}
\fontsize{10}{10}\selectfont
\begin{tabular}{|c|c|c|c|c|c|c|c|c|}
\hline \multirow{2}{*}{$POT$}& \multirow{2}{*}{$\rho(fm^{-3})$}&
\multicolumn{2}{c|}{$U_{PNM}^{T=1}(MeV)$}&\multicolumn{2}{c|}{$U_{SNM}^{T=1}(MeV)$}&\multicolumn{2}{c|}{$U_{SNM}^{T=0}(MeV)$}&
\multirow{2}{*}{$E_{sym}(MeV)$}\cr\cline{3-8}&¡¡&$L=odd$&$L=even$&$L=odd$&$L=even$&$L=odd$&$L=even$&\cr
\hline \hline \multirow{2}{*}{$AV18$}& 0.175 & -0.253 & -23.437  &
-0.341  & -19.981 & 5.046 & -24.859 & 30.038 \cr\cline{2-9}& 0.400
& 11.739 & -39.210  &  3.219  & -36.342 & 13.620& -35.877 &
50.866\cr \hline \multirow{2}{*}{$BONN\ B$}& 0.175  & 0.295 & -23.414
& -0.136  & -20.438 & 5.770 & -25.458 & 30.230
\cr\cline{2-9}&0.400 & 10.105 & -39.626 & 2.948  & -37.170 &
16.218 & -39.066 & 50.470\cr \hline
\end{tabular}
\end{center}
\caption{ Partial wave decomposition of the potential
energy from BHF with 2BF.}
\end{table}
 It amounts to say that the spin-singlet contribution is
vanishing due to the generalized Pauli principle ($L+S+T=odd$), when
restricting to only $L=0$ angular momentum. The conclusion is that
the spin-triplet two-body channel $^3S_1$ gives the largest
contribution to the symmetry potential energy, being $L > 0 $
channels much smaller. This result was already found in the
earliest $ab \, initio$ calculations of the symmetry
energy\cite{naka,bomb}. At supra-saturation density the isospin
symmetry is  violated so that the isospin-singlet starts to compete
with isospin-triplet. The interplay between isospin-singlet and
isospin-triplet is quite clearly illustrated in Fig.2. The
decomposition of $T=0$ and $T=1$ potential energy per particle is
reported in Table I for two densities. It is worthwhile noticing
the rather good agreement between the two interactions adopted in
the BHF calculations, giving comparable values for the symmetry
energy as well as the the individual components, except for the
$T=0$ ones at the higher density. At saturation density the
isospin-triplet components slightly change going from SNM to PNM,
leaving the isospin-singlet T=0 term to play the major role in
determining the symmetry energy, as discussed before. At the
higher density also the isospin-triplet T=1 to contribute to the
enhancement of the symmetry energy.
\begin{figure}[htb]
\centering
\includegraphics[angle=0,width=12cm]{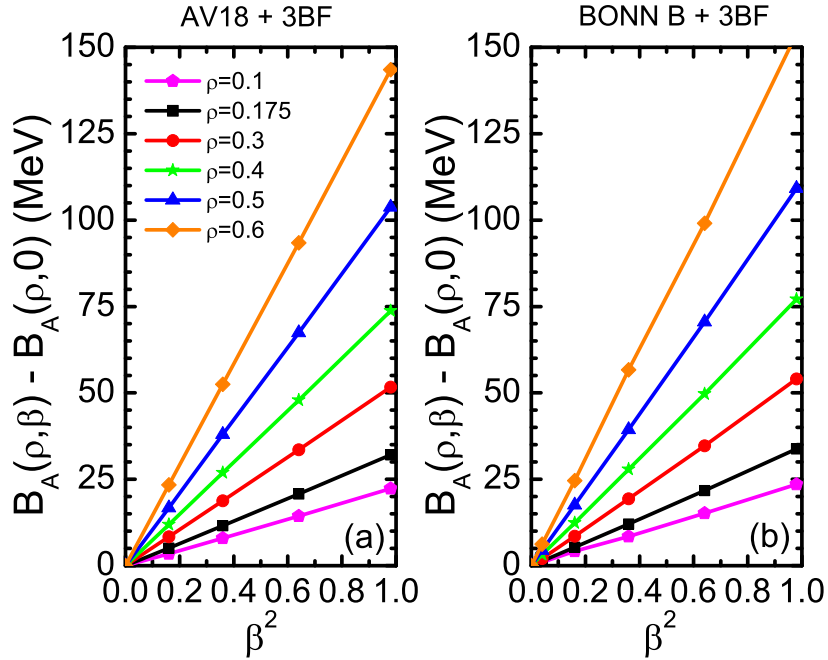}
\caption{Energy per particle in ANM vs. $\beta^2$, including 3BF
in the two versions: consistent with AV18 (left), and consistent
with Bonn B (right).}\label{anm3}
\end{figure}
As already firmly established the 3BF is necessary to reproduce
the saturation density of nuclear matter. At super-saturation
density the 3BF  becomes the dominant interaction. The simplest
way to extend the BHF approximation is to replace the 3BF by a
density dependent 2BF weighting the effect of the third particle
by means of its correlation with the other two particles
\cite{av3bf}. In coordinate space it can be written formally
\begin{equation}
W(r_{12})\,=\,\rho \int d^3r_3  V(r_1,r_2,r_3) g^2(r_{13})
g^2(r_{23})
\end{equation}
where $g(r)=1-\eta(r)$, $\eta(r)$ being the defect function. The
energy per particle in ANM is still a $\beta^2$ function for all
densities considered, as shown in Fig.3. The 3BF contribution to
the isospin shift of the single-particle potential $\Delta
u_{\tau}$ is given by
\begin{equation}
\Delta u_n -\Delta u_p  \,=\, \frac{1}{2}\rho [\tilde\mathcal{
W}_{nn}+\tilde\mathcal{W}_{pp}-2\tilde\mathcal{ W}_{np}](n_n+n_p)
\beta,
\end{equation}
where
\begin{eqnarray}
\nonumber
\tilde\mathcal{W}_{\tau\tau}&=& V_{\tau\tau\tau}
g^2_{\tau\tau}+
V_{\tau\tau\tau'} g^2_{\tau\tau'} \\
\tilde\mathcal{W}_{\tau\tau'}&=& V_{\tau\tau'\tau'}
g_{\tau\tau'}g_{\tau'\tau'}+ V_{\tau\tau'\tau}
g_{\tau\tau}g_{\tau'\tau'},
\end{eqnarray}
and $\tau\neq\tau'$. This contribution is weakly asymmetry
dependent, the same as the 2BF term, so that the
$\beta^2$-linearity is to be expected. The numerical results in
fact confirm such a property, as shown in Fig.3, for both 3BF:
meson-exchange 3BF consistent with Argonne V18 \cite{3bf1} and
meson-exchange 3BF consistent with Bonn B \cite{3bf2}.
\begin{figure}[htb]
\centering
\includegraphics[angle=0,width=12cm]{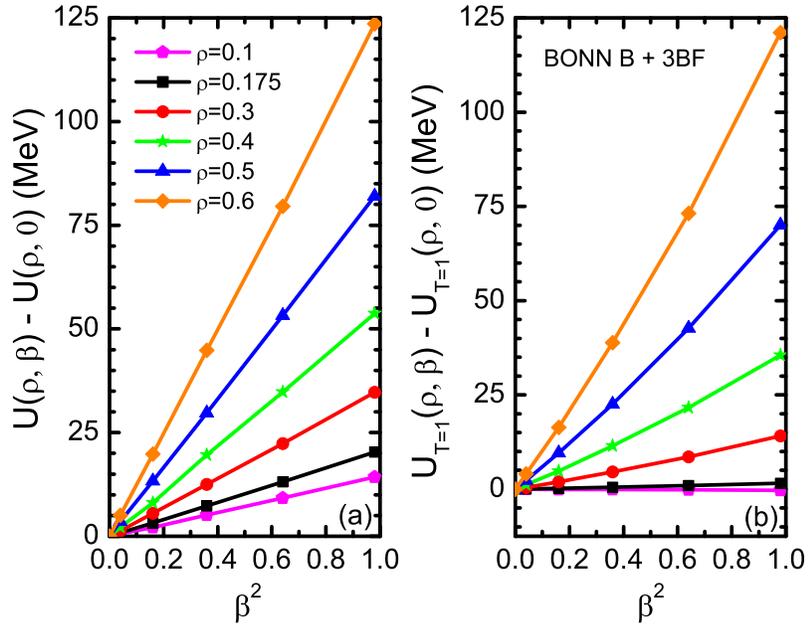}
\caption{Total potential energy per particle in ANM vs.
$\beta^2$(left), and isospin triplet contribution (right).
}\label{anm3}
\end{figure}
Now the interplay between isospin singlet and triplet is displayed
in Fig.4. It is dominated by the 3BF, whose strength is strongly
increasing with density. The isospin-triplet contribution reaches
$95\%$ of the total symmetry potential energy at the highest
considered density $\rho = 0.6 fm^{-3}$.

The decomposition of isospin singlet and triplet in L partial
waves is reported in Table II. In this case the agreement between
the two 3BF adopted in the calculations is not so good as before,
but the difference in the symmetry energy is about $5\%$.
Since the 3BF strength is still small at the saturation density,
the isospin-triplet is dominated by the isospin-singlet, but at
higher density it participates (with odd L) to increase the
symmetry energy at the same footing as the isospin-singlet (with
even L), as noticed in Fig.4. In conclusion the 3BF reinforces the
spin-triplet component of the full interaction.
\begin{table}
\renewcommand{\arraystretch}{1.5}
\begin{center}
\fontsize{10}{10}\selectfont
\begin{tabular}{|c|c|c|c|c|c|c|c|c|}
\hline \multirow{2}{*}{$POT$}& \multirow{2}{*}{$\rho(fm^{-3})$}&
\multicolumn{2}{c|}{$U_{PNM}^{T=1}(MeV)$}&\multicolumn{2}{c|}{$U_{SNM}^{T=1}(MeV)$}&\multicolumn{2}{c|}{$U_{SNM}^{T=0}(MeV)$}&
\multirow{2}{*}{$E_{sym}(MeV)$}\cr\cline{3-8}&¡¡&$L=odd$&$L=even$&$L=odd$&$L=even$&$L=odd$&$L=even$&\cr
\hline \hline $AV18$ & 0.175 & -1.752 & -17.522  & -2.707  & -15.791 & 3.570 & -23.335 & 32.092\cr\cline{2-9}
+$\ 3BF$ &0.400 & 17.467 & -6.626  &  -4.146  & -15.159 & 7.390 & -27.875 & 73.758\cr
\hline
$BONN\ B$& 0.175  & -2.670 & -15.509 & -3.194  & -16.568 & 2.398 & -21.510 & 33.800\cr\cline{2-9}
+$\ 3BF$ &0.400 & 14.260 & 6.487 & -6.040  & -8.772 & 2.188 & -20.578 & 77.110\cr \hline
\end{tabular}
\end{center}
\caption{Partial wave decomposition of the potential
energy from BHF with 3BF.}
\end{table}

In the present note some properties of nuclear symmetry energy,
calculated within the BHF approximation with two and three body
forces, have been discussed in connection with symmetries of the
interaction. The spin-isospin decomposition was performed to
illustrate the interplay between different two body channels in
terms of isospin singlet and triplet components as well as the
spin singlet and triplet ones. This investigation is a preliminary
step to calculate the nuclear symmetry energy beyond the mean
field approximation, including medium polarization effects.

\begin{acknowledgments}
The authors thank Dr. Z. H. Li for providing the BHF code with Bonn
B and 3BF and Dr. M. Baldo, and Dr. I. Vida$\tilde{n}$a for valuable
discussions. This work was supported by INFN fellowship of Italy and the National Natural Science
Foundation of China under Grants No. 11705109. 
\end{acknowledgments}

\end{document}